\documentclass[a4paper,11pt]{article}
\usepackage{jcappub}
\usepackage[T1]{fontenc}

\usepackage{comment}
\usepackage[table,dvipsnames]{xcolor}
\usepackage{enumitem}
\usepackage{tikz}
\usepackage{multirow}
\usepackage{bbm}
\DeclareMathAlphabet{\mathbbmsl}{U}{bbm}{m}{sl}

\newcommand{\fsampHz}{17}
\newcommand{\slopef}{{\alpha}}
\newcommand{\slopel}{{\beta}}
\newcommand{\alphascan}{{\alpha}}
\newcommand{\betascan}{{\beta}}
\newcommand{\wnl}{{W}}
\newcommand{\offsetarcmin}{{50}}
\newcommand{\doubleoffsetarcmin}{{100}}

\newcommand{\minor}[1]{}

\title{Efficient GLS map-making and assessment of $1/f$ noise impact\\ for future CMB satellite missions}
\author[a]{Marta Monelli,}
\author[b,c,a]{Guillaume Patanchon,}
\author[d,e]{and Avinash Anand.}

\affiliation[a]{
Kavli Institute for the Physics and Mathematics of the Universe (Kavli IPMU, WPI), UTIAS, The University of Tokyo, Chiba, 277-8583, Japan
}
\affiliation[b]{ILANCE, CNRS – University of Tokyo International Research Laboratory, Kashiwa, Chiba 277-8582, Japan}
\affiliation[c]{Université Paris Cité, F-75006 Paris, France}
\affiliation[d]{Dipartimento di Fisica, Università di Roma Tor Vergata, Via della Ricerca Scientifica, 1, 00133, Roma, Italy}
\affiliation[e]{INFN Sezione di Roma2, Università di Roma Tor Vergata, Via della Ricerca Scientifica, 1, 00133, Roma, Italy}

\emailAdd{marta.monelli@ipmu.jp}

\abstract{
Time-correlated \(1/f\) noise is a major source of large-scale contamination in measurements of the polarized cosmic microwave background (CMB) that can limit sensitivity to primordial \(B\) modes. Generalized least-squares (GLS) map-making is optimal for stationary Gaussian noise when its covariance is known, but practical implementations rely on approximations to make the computation tractable. In particular, approximating the noise covariance as circulant enables efficient map-making using fast Fourier transform, but introduces artificial edge effects when applied to finite data chunks.

We introduce STITCH (STatistical Inpainting for Timeline CHunks), a preprocessing method that mitigates these effects by extending time-ordered data chunks with synthetic samples that make the extended timelines statistically consistent with the assumed circulant noise model. They can then be processed with standard circulant GLS solvers, preserving their computational efficiency while reducing residuals associated with chunk boundaries. Simulations show that STITCH substantially suppresses these edge effects, bringing the resulting noise power spectra close to those obtained when the circulant approximation holds. We demonstrate negligible impact of our procedure on the recovered signal, such as $E\to B$ leakage, even at large scales.

Finally, we investigate how temporal \(1/f\) noise projects onto angular scales for representative LiteBIRD-, \emph{Planck}-, and PICO-like scanning strategies. Using noise simulations spanning a range of temporal noise parameters, we fit an effective model to the resulting angular noise spectra. These results provide an empirical characterization of the temporal-to-angular noise mapping for the configurations studied and a basis for further investigations of correlated noise in future CMB polarization missions.
}

\begin{document}

\maketitle

\section{Introduction}
Since its discovery, observations of the cosmic microwave background (CMB) have provided increasingly precise constraints on cosmological models and the physics of the early Universe~\cite{WMAP:2012,Planck:2018cosmpar,AtacamaCosmologyTelescope:2025blo}. Current and upcoming CMB experiments~\cite{Ade_2019, Moncelsi:2020ppj, LiteBIRD:2022cnt, LiteBIRD:2026gdf, NASAPICO:2019thw} aim to detect the faint primordial $B$ modes, which would provide evidence for primordial gravitational waves and inflationary physics~\cite{Seljak:1996gy,Kamionkowski:1996zd} through a measurement of the tensor-to-scalar ratio, $r$.

The inflationary $B$-mode signal is expected to be strongest on large angular scales, making space missions such as the upcoming LiteBIRD~\cite{LiteBIRD:2022cnt, LiteBIRD:2026gdf} and the proposed PICO~\cite{NASAPICO:2019thw} particularly well suited for measuring $r$. As they are not affected by atmospheric emission and can observe the full sky, they can achieve unprecedented sensitivity to this parameter.

As experimental sensitivity improves, the mitigation of instrumental and astrophysical systematics becomes increasingly critical. One of the dominant sources of uncertainty in CMB experiments is time-correlated $1/f$ noise, characterized by higher power at low frequencies. If unmitigated, $1/f$ noise projects onto the observed maps as a strong contaminant on large angular scales and can easily obscure the primordial $B$ modes at low multipoles. 

We can identify three complementary classes of approaches to mitigate the impact of time-correlated noise in CMB observations:
\begin{description}[topsep=2mm,parsep=1mm,itemsep=1mm,leftmargin=6mm]
    \item[Polarization modulation.] A rapidly rotating half-wave plate (HWP) as first optical element modulates the polarized sky signal to high temporal frequencies, away from the $1/f$ slope \cite{ABS:2013dqh}. This is an effective mitigation strategy at the hardware level, but comes with technical challenges (especially for space missions) and inevitably introduces additional systematic effects that must be carefully characterized and controlled~\cite{Verges:2020xug, Duivenvoorden:2020xzm, Giardiello:2021uxq, Monelli:2022pru, Patanchon:2023ptm, Monelli:2023wmv,
    Micheli:2024hfe, LiteBIRD:2026qdz, TsangKingSang:2026rkn}.
    \item[Scanning strategy.] By relating temporal frequencies to angular scales on the sky, the scanning strategy strongly influences how correlated noise projects onto the observed maps.
    \item[Map-making.] Finally, the choice of map-making method \cite{Tegmark:1996qs} determines how detector timelines are processed into sky maps and therefore how correlated noise projects.
\end{description}
In this work we focus on the latter two approaches. 

We introduce STITCH (STatistical Inpainting for Timeline CHunks), a preprocessing technique that synthetically extends time-ordered data chunks, so that they satisfy periodic boundary conditions while remaining statistically consistent with the assumed stationary noise model. This allows us to efficiently process the extended chunks with circulant Generalized Least-Squares (GLS) map-making, producing near-optimal maps. 

We then use this framework to quantify how different scanning strategies project temporal $1/f$ noise onto the observed maps. Considering representative LiteBIRD-, \emph{Planck}-, and PICO-like satellite missions, we derive empirical relations between the temporal noise spectrum and its angular counterpart.

The remainder of this paper is organized as follows. In section~\ref{sec:math}, we introduce the map-making problem in the presence of time-correlated noise and discuss the commonly adopted circulant approximation, together with its limitations. In section~\ref{sec:stitch}, we introduce a novel map-making method that combines an inpainting preprocessing step with circulant GLS and discuss its numerical implementation in the map-making framework \texttt{BrahMap}\footnote{\url{https://github.com/anand-avinash/BrahMap}} \cite{Anand:2025qio}. We show that the method introduces negligible bias in the recovered signal. In section~\ref{sec:scanstr}, we apply our tool to simulations for different choices of noise parameters and satellite scanning strategies. We then use the resulting angular power spectra to derive empirical relations between temporal and angular noise parameters. Section~\ref{sec:validation} presents a detailed validation of the method, while the specific choice of method parameters is discussed in appendix~\ref{subsec:params}. We summarize our findings and conclusions in section~\ref{sec:conclusions}.

\section{The map-making problem with $1/f$ noise}\label{sec:math}
In a CMB experiment, $n_\text{det}$ detectors collect $n_\text{obs}$ samples each over the observation period. Together, these $n_\text{det}n_\text{obs}$ samples form the time-ordered data (TOD). Assuming a linear instrument response and neglecting asymmetries in the beam, we can model the TOD, $\mathbf{d}$, as
\begin{equation}\label{eqn:data_model_TOD}
    \mathbf{d} = A \cdot \mathbf{m}_\text{in} + \mathbf{n}\,,
\end{equation}
where $\mathbf{m}_\text{in}$ denotes the beam-smoothed Stokes $\{I,Q,U\}$ pixelized sky maps\footnote{No mechanism proposed to source circular polarization within or beyond the standard cosmological model (see, for instance, \cite{Cooray:2002nm,Alexander:2008fp,Bavarsad:2009hm,Sadegh:2017rnr,Vahedi:2018abn,Inomata:2018vbu,Bartolo:2019eac,Alexander:2019sqb,Lembo:2020ufn}) allows for a significant signal. We therefore neglect it.}; $A$ is the response matrix, encoding the telescope pointings and the instrument response; and $\mathbf{n}$ represents the noise contribution. 
The response matrix $A$ has dimensions $n_{\rm det}n_{\rm obs}\times3n_{\rm pix}$, with one row for each TOD sample and three columns associated to each pixel (one per Stokes parameter $\{I,Q,U\}$). The $(i\, n_{\rm obs}+t)$ row of $A$ contains only three non-zero elements corresponding to the pixel $p_{it}$ observed by detector $i$ at time $t$, which we can represent as the row vector $\mathbbmsl{S}^T_{it}$. For instance, assuming the detector $i$ to be a perfect polarimeter located at the telescope boresight, its instrumental response can be modeled as
\begin{equation}
    \mathbbmsl{S}^T_{it} = \begin{pmatrix}
    1 & \cos[2(\psi_t+\xi_i)] & \sin[2(\psi_t+\xi_i)]
    \end{pmatrix},
\end{equation}
where $\psi_t$ is the angle between the telescope and sky coordinate systems at time $t$, and $\xi_i$ is the polarization orientation of the detector.

Pixelized representation of the sky is another inherent assumption encoded in the response matrix of eq.~\eqref{eqn:data_model_TOD}. 
In realistic observations, detector pointings can fall anywhere within a pixel, whereas the model represents the sky using a finite set of pixel values. This mismatch gives rise to sub-pixel effects, which can introduce additional residuals in the reconstructed polarization maps, largely coming from Stokes parameter mixing. We initially neglect these effects and examine them separately in section \ref{subsec:subpix}.

\paragraph{Characterizing the noise}
As for the noise term, \(\mathbf{n}\), various effects can contribute to it, and accurately modeling them is beyond the scope of this work. Here, we assume the noise to be Gaussian and stationary, in which case its statistical properties are fully characterized by its power spectrum, \(P(f)\), or, equivalently, by its continuous-time covariance function, $R(\tau) = \mathcal{F}^{-1} \left[P(f)\right](\tau)$, where \(\tau\) denotes the time lag and $\mathcal{F}$ the Fourier transform.

\begin{figure}[!t]
\centering
\includegraphics[trim={3mm 3mm 0 3mm},clip]{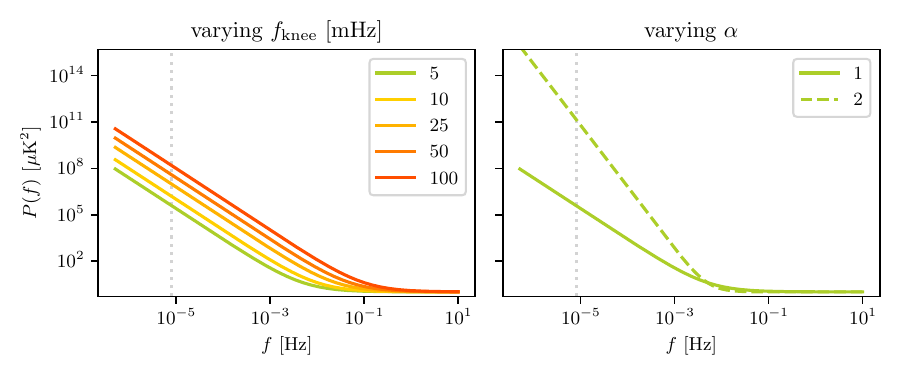}
\caption{Noise power spectrum, \(P(f)\), as in eq.~\eqref{eqn:Pf}, for different choices of the model parameters. In the left panel, we vary \(f_\text{knee}\) while keeping \(\slopef=1\). In the right panel, we vary \(\slopef\) at fixed \(f_\text{knee}=5\) mHz. The white-noise power level is normalized to unity. We show the frequency range from the minimum correlation frequency considered in this work (corresponding to \(\sim23\) days) to \(f_\text{Nyquist} \equiv f_\text{samp}/2\). The vertical dotted line indicates the frequency corresponding to the \(\sim1.2\) day chunks.}
\label{fig:noise_Pf}
\end{figure}

In CMB experiments, the noise is typically modeled as a superposition of white and \(1/f\) components \cite{Ihle:2020jcq}:
\begin{equation}\label{eqn:Pf}
P(f) = \sigma^2 \left[1+\left(\frac{f_\text{knee}}{f}\right)^\slopef\right],
\end{equation}
where \(\sigma\) quantifies the white-noise amplitude; the knee frequency, \(f_\text{knee}\), represents the frequency at which the \(1/f\) and white-noise components contribute equally to the power spectrum; and \(\slopef\) is the slope of the correlated component. To account for finite stationary periods and avoid the singularity at \(f=0\), we introduce an additional parameter \(f_\text{min}\):
\begin{equation}\label{eqn:Pf_fmin}
P(f) = \sigma^2 \left(\frac{f^\slopef + f_\text{knee}^\slopef}{f^\slopef + f_\text{min}^\slopef}\right).
\end{equation}
Figure~\ref{fig:noise_Pf} illustrates the resulting spectrum: at high frequencies, the spectrum approaches the white-noise floor, while at low frequencies the \(1/f^\slopef\) rise dominates, with \(f_\text{knee}\) marking the transition between the two regimes. The values of \(f_\text{knee}\) and \(\slopef\) shown in the plot (\(f_\text{knee}\in \{5,10,25,50,100\}\,\mathrm{mHz}\) and \(\slopef\in\{1,2\}\)) are motivated by \emph{Planck} performance~\cite{2011A&A...536A...6P, Ihle:2020jcq} and are used as reference values throughout this paper. The effect of \(f_\text{min}\) is not visible because we choose \(f_\text{min}=10^{-8}\,\mathrm{Hz}\) throughout this paper, below the frequency range shown.

For a discretely sampled timestream with sampling interval \(\Delta t\), the noise covariance matrix is $N \equiv \left\langle \mathbf{n}\mathbf{n}^{T}\right\rangle$. If the noise is stationary, \(N\) is a symmetric Toeplitz matrix with elements
\begin{equation}\label{eqn:Toeplitz}
N_{ij} = R \left[(i-j)\Delta t\right].
\end{equation}
If we additionally impose periodic boundary conditions on the timestream, the corresponding covariance matrix becomes circulant. Denoting a circulant covariance matrix by \(C\), its elements can be written in terms of the elements of its first column, $c$, as
\begin{equation}
C_{ij} = c\left({(i-j)\bmod n_\text{samp}}\right)\,.
\end{equation}

\subsection{GLS map-making}\label{subsec:GLS}
The map-making step translates the TOD into a set of pixelized maps, $\widehat{\mathbf{m}}$, with the goal of recovering the underlying sky maps, $\mathbf{m}_\text{in}$. Among the several different methods that have been proposed in the past decades \cite{Tegmark:1996qs}, generalized least-squares (GLS) map-making provides an optimal framework for handling time-correlated noise
\begin{equation}\label{eqn:mapmaker}
    \widehat{\mathbf{m}} = \left(\widehat{A}^T \widehat{N}^{-1} \widehat{A}\right)^{-1} \widehat{A}^T \widehat{N}^{-1} \mathbf{d}\,,
\end{equation}
where $\widehat{A}$ and $\widehat{N}$ are the assumed response and noise covariance matrices, respectively. If the assumed response and noise covariance matrices, $\widehat{A}$ and $\widehat{N}$, coincide with the true ones, the GLS solution is the maximum-likelihood estimate of the sky map\footnote{For a given noise realization $\mathbf{n}$, we define its whitened counterpart as $\mathbf{n}' \equiv W\mathbf{n}$, where the whitening matrix $W$ is chosen such that $W N W^T = \mathbbm{1}$. In particular, this condition holds if we choose $W$ to be the symmetric square root of $N^{-1}$: $W^T W=N^{-1}$. Multiplying the data model in eq.~\eqref{eqn:data_model_TOD} by $W$ then gives $W\mathbf{d} = WA,{\mathbf{m}}_\text{in} + W\mathbf{n}$, where $W\mathbf{n}$ has unit covariance. The map can therefore be estimated by ordinary least squares applied to the whitened data:
\begin{equation*}
\widehat{\mathbf{m}}
= \left[(WA)^T(WA)\right]^{-1}(WA)^T W\mathbf{d}
= \left(A^T W^T W A\right)^{-1}A^T W^T W\mathbf{d}
= \left(A^T N^{-1}A\right)^{-1}A^T N^{-1}\mathbf{d}.
\end{equation*}
} \cite{Tegmark:1996qs}. 

The GLS solution has been widely used in CMB map-making, with practical implementations adopting different numerical approaches and assumptions about the noise covariance \cite{Natoli:2001tb, Dore:2001nf, Stompor:2002jy, deGasperis:2005di, Yvon:2004pb, Patanchon:2007yw, 2010ApJS..187..212C, 2013ApJ...762...10D}. A common approach is to rewrite Eq.~\eqref{eqn:mapmaker} as
\begin{equation}\label{eqn:mapmakerCG}
    \left(\widehat{A}^T \widehat{N}^{-1} \widehat{A}\right) \, \widehat{\mathbf{m}} = \widehat{A}^T \widehat{N}^{-1} \mathbf{d}\,,
\end{equation}
and solve it numerically using the conjugate gradient method \cite{Shewchuk1994AnIT}. This avoids explicitly inverting $\widehat{A}^T \widehat{N}^{-1} \widehat{A}$, but each iteration still requires applying $\widehat{N}^{-1}$ to a TOD-like vector. Since $\widehat{N}$ is an $n_\text{det}n_\text{obs} \times n_\text{det}n_\text{obs}$ matrix, explicitly constructing or inverting a dense noise covariance matrix can be computationally prohibitive for modern CMB experiments. Practical GLS implementations therefore rely on simplifying assumptions about the noise covariance that allow $\widehat{N}^{-1}$ to be applied efficiently without constructing the full matrix.

\subsection{The circulant approximation and its limitations: edge effects}\label{subsec:edge}
{For computational feasibility, GLS map-makers typically process the TOD in finite chunks rather than as a single continuous timeline. Chunk boundaries may coincide with natural gaps in the data, such as those caused by data transfer or observing-system operations, or be introduced when no such gaps are present. Within each chunk, the noise is approximated as stationary, so that the covariance depends only on the time lag between samples, as in eq.~\eqref{eqn:Toeplitz}. The resulting covariance matrix is Toeplitz within each chunk.}

A further approximation is commonly introduced by imposing periodic boundary conditions on each chunk. As discussed above, this makes the covariance matrix circulant. This approximation is particularly useful computationally because circulant matrices are diagonalized by the discrete Fourier transform (DFT). Denoting the assumed circulant covariance matrix by \(C\), we can write
\begin{equation}
C = F^{-1} P_f F,
\end{equation}
where \(F\) denotes the DFT matrix and \(P_f\) is a diagonal matrix with diagonal elements given by the discrete noise power spectrum associated with the finite, sampled timestream. Its inverse can therefore be applied to a vector \(\mathbf{v}\) as
\begin{equation}\label{eqn:fourierspaceC}
C^{-1}\mathbf{v}
=
\operatorname{IDFT}
\left[
P_f^{-1}\operatorname{DFT}(\mathbf{v})
\right].
\end{equation}
Thus, applying the inverse noise covariance reduces to two Fourier transforms and a element-wise vector multiplication in Fourier space, avoiding the explicit construction and inversion of a dense covariance matrix and substantially reducing the computational complexity.

Although these assumptions make eq.~\eqref{eqn:mapmakerCG} tractable, they are only approximations. Real observations inevitably contain data gaps, flagged samples, and other interruptions, which complicate the direct application of a stationary covariance model. In addition, the periodic-boundary assumption is not physically justified: the long-timescale drifts characteristic of $1/f$ noise imply that the beginning and end of a data segment are generally unrelated. In either case, the circulant approximation does not faithfully represent the true covariance matrix, and the resulting mismatch propagates into the reconstructed maps as a large-scale noise residual.
In this work we focus on the \emph{edge effects} induced by the unphysical periodic boundary conditions imposed by the circulant approximation. In section~\ref{subsec:statistics}, we comment on how the framework developed here could be extended to account for gaps in the data. 

\begin{figure}[!t]
    \centering
    \includegraphics{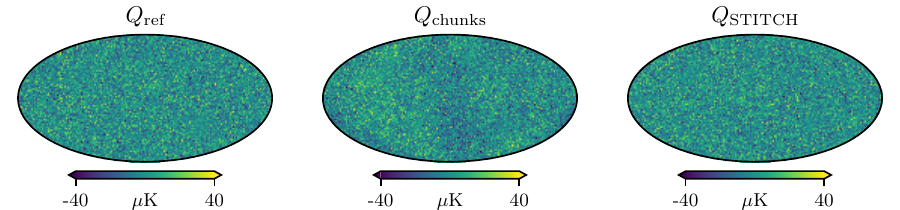}
    \caption{Stokes $Q$ maps obtained from a $1/f$ noise realization which is circulant over $\sim 23$ day segments. The differences in the maps depend on the choice of map-making method. The reference solution (left) uses a circulant GLS over the full circulant segments. The center and right panels are instead obtained by processing the same noise realization over $\sim 1.2$ day stationary, but not circulant, chunks. The large-scale features visible when using the circulant GLS (center) disappear when the TOD is preprocessed with STITCH (right). For the specifics of the simulation, see section~\ref{subsec:noise}.}
    \label{fig:nostat}
\end{figure}

Figure~\ref{fig:nostat} illustrates the impact of edge effects by comparing the Stokes $Q$ maps obtained by processing the same noise timeline with a circulant GLS map-maker that operates on segments that satisfy the periodic boundary condition ($Q_{\rm ref}$, left panel) and on shorter stationary, but not circulant, chunks ($Q_{\rm chunks}$, center panel). Although the underlying noise realization is identical, $Q_{\rm chunks}$ shows large-scale residuals that are absent in $Q_{\rm ref}$. 
These residual modes substantially increase the recovered angular noise power spectrum at low multipoles. As a preview of figure~\ref{fig:cls_together}, maps obtained by processing circulant segments in shorter chunks exhibit more than two orders of magnitude larger noise power at low multipoles than those obtained from the full segments, which can significantly degrade the sensitivity of the satellite missions to the tensor-to-scalar ratio $r$.

\section{Mitigating edge effects with STITCH}\label{sec:stitch}
The circulant assumption imposes artificial periodic boundaries on each TOD chunk, which causes edge effects, as shown in section~\ref{sec:math}. Rather than modifying the circulant GLS map-making procedure, we preprocess the input timeline to mitigate these effects. The key idea is to extend each chunk with synthetic samples that restore the long-timescale correlations missing at its boundaries. These samples are not intended to represent additional observations: they provide a statistically motivated extension of the timeline that makes the circulant approximation {hold}. 

We implement this approach in STITCH (STatistical Inpainting for Timeline CHunks), which estimates the synthetic extension from the observed samples and the assumed noise covariance. To incorporate the synthetic samples into the map-making equations, we assign them to auxiliary pixels that are excluded from the final maps. 
The extended data and pointing timelines can then be processed using standard circulant GLS machinery, retaining its computational advantages while reducing edge effects. 


\minor{In this section, we introduce the inpainting method, describe its implementation in the circulant GLS map-making pipeline, and evaluate its impact on noise residuals and signal reconstruction.}

\subsection{Statistical formulation of the inpainting problem}\label{subsec:statistics}
Our goal is to reconstruct a synthetic extension of the observed TOD chunk such that the resulting timeline is statistically consistent with the assumed circulant noise model. We formulate the boundary restoration as a conditional Gaussian inference problem. 
Constrained Gaussian realizations and conditional inferences have a rich history in cosmology and CMB data processing. The underlying formalism was first proposed in the context of $N$-body simulations \cite{1991ApJ...380L...5H} and later applied in CMB data analysis to fill observational gaps in the timelines (e.g., \cite{Stompor:2002jy}) as well as masked regions in maps (e.g., \cite{2012MNRAS.424.1694B}). Here, we use this formalism specifically to generate artificial timeline extensions at chunk boundaries, {eliminating the edge effects induced by processing non-circulant data chunks with circulant GLS.}



Given a noise-only TOD chunk, $\mathbf{n}$, of length $n_\text{samp}$, we want to construct a synthetic extension, $\mathbf{x}$, of length $n_\text{ext}$, such that the concatenated TOD $[\mathbf{n},\mathbf{x}]$ of length $n_\text{tot} \equiv n_\text{samp} + n_\text{ext}$ can be treated as a realization of a stationary Gaussian process with circulant covariance, $C$. {The covariance matrix $C$ is constructed from the $1/f$ noise power spectrum, $P_f$, specified by the assumed noise parameters, for an extended timeline containing $n_\mathrm{tot}$. Specifically, 
we require that the concatenated timeline} $[\mathbf{n},\mathbf{x}]$ follows a multivariate Gaussian distribution with zero mean:
\begin{equation}\label{eqn:mvg}
    p(\mathbf{n}, \mathbf{x}|C) = \frac{1}{(2\pi)^{n_\text{tot}/2}\sqrt{\vert C\vert}}\exp\left(-\frac{1}{2}[\mathbf{n}, \mathbf{x}]^T C^{-1} [\mathbf{n}, \mathbf{x}]\right),
\end{equation}
where $\vert C\vert$ is the determinant of $C$. {We can then write} the probability distribution function for $\mathbf{x}$, given $\mathbf{n}$ and $C$, as\footnote{
According to the definition of conditional probability, 
\begin{equation*}
    p(\mathbf{x}\,\vert\, \mathbf{n},C) = \frac{p(\mathbf{n},\mathbf{x},C)}{p(\mathbf{n},C)} = \frac{p(\mathbf{n},\mathbf{x}\,\vert\, C)}{p(C)}\frac{p(C)}{p(\mathbf{n}\,\vert\, C)} = \frac{p(\mathbf{n},\mathbf{x}\,\vert\, C)}{p(\mathbf{n}\,\vert\, C)}\,.
\end{equation*}
}
\begin{equation}\label{eqn:P_gauss}
    p(\mathbf{x}\,\vert\, \mathbf{n},C) = \frac{1}{p(\mathbf{n}\,\vert\, C)} \frac{1}{(2\pi)^{n_\text{tot}/2}\sqrt{\vert C\vert}}\exp\left(-\frac{1}{2}[\mathbf{n}, \mathbf{x}]^T C^{-1} [\mathbf{n}, \mathbf{x}]\right).
\end{equation}
The maximum of the conditional distribution in eq.~\eqref{eqn:P_gauss}, which we denote by $\overline{\mathbf{x}}$, satisfies
\begin{equation}\label{eqn:condition}
    \frac{\partial p(\mathbf{x}\,\vert\, \mathbf{n},C)}{\partial \mathbf{x}}\biggr\vert_{\overline{\mathbf{x}}} = 0\,.
\end{equation}
Using eq.~\eqref{eqn:P_gauss}, we rewrite the derivative as
\begin{align}
    \frac{\partial p(\mathbf{x}\,\vert\, \mathbf{n},C)}{\partial \mathbf{x}} &\propto \frac{\partial}{\partial \mathbf{x}}\left([\mathbf{n}, \mathbf{x}]^T C^{-1} [\mathbf{n}, \mathbf{x}] \right)\nonumber\\
    &\propto \frac{\partial}{\partial \mathbf{x}}\left[
    \mathbf{n}^T \left(C^{-1}\right)_{nn} \mathbf{n} + 
    \mathbf{n}^T \left(C^{-1}\right)_{nx} \mathbf{x} +
    \mathbf{x}^T \left(C^{-1}\right)_{xn} \mathbf{n} + 
    \mathbf{x}^T \left(C^{-1}\right)_{xx} \mathbf{x}
    \right],
\end{align}
where the $nn$, $nx$, $xn$ and $xx$ subscripts denote the different blocks of $C^{-1}$. By expanding the derivatives and using the symmetry of $C^{-1}$, we write
\begin{equation}
    \frac{\partial p(\mathbf{x}\,\vert\, \mathbf{n},C)}{\partial \mathbf{x}} \propto \left(C^{-1}\right)_{xn} \mathbf{n} + \left(C^{-1}\right)_{xx} \mathbf{x}\,,
\end{equation}
which allows us to express eq.~\eqref{eqn:condition} as $\left(C^{-1}\right)_{xx}\overline{\mathbf{x}} = -\left(C^{-1}\right)_{xn}\mathbf{n}$. Equivalently, in terms of embedding operators, we can write
\begin{equation}\label{eqn:realspace}
    (\mathcal{P}_x^T C^{-1} \mathcal{P}_x)\, \overline{\mathbf{x}} = - (\mathcal{P}_x^T C^{-1} \mathcal{P}_n)\, \mathbf{n}\,,
\end{equation}
{where $\mathcal{P}_n$ and $\mathcal{P}_x$ are embedding operators that map a vector of size $n_\mathrm{samp}$ and $n_\mathrm{ext}$, respectively, into a vector of size $n_\mathrm{tot} = n_\mathrm{samp} + n_\mathrm{ext}$.}
{In a toy example with $n_\text{samp}=3$ and $n_\text{ext}=2$, $\mathcal{P}_n$ and $\mathcal{P}_x$ read
\begin{equation}
    \mathcal{P}_n = 
    \begin{pmatrix}
        1 & 0 & 0 \\
        0 & 1 & 0 \\
        0 & 0 & 1 \\
        0 & 0 & 0 \\
        0 & 0 & 0 
    \end{pmatrix},
    \qquad 
    \mathcal{P}_x = 
    \begin{pmatrix}
        0 & 0 \\
        0 & 0 \\
        0 & 0 \\
        1 & 0 \\
        0 & 1 
    \end{pmatrix}.
\end{equation}
This embedding-operator formulation is useful because it does not depend on the synthetic samples being appended to the observed timeline. By using different embedding operators, eq.~\eqref{eqn:realspace} can be used to reconstruct arbitrary gaps or flagged samples.}

\subsection{Numerical implementation}\label{subsec:num_imp}

\begin{figure}
\begin{minipage}{0.47\textwidth}
    \centering
    \includegraphics[trim={1.75mm 2mm 0 0},clip]{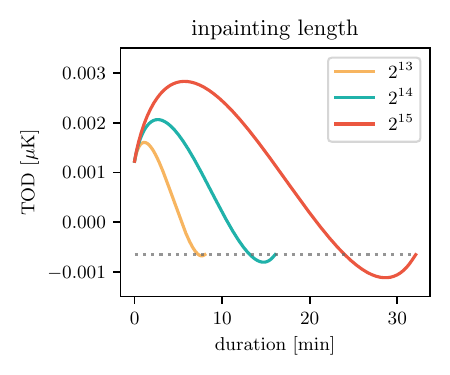}
    \caption{Synthetic extensions for a few different choices of the inpainting length, $n_\mathrm{ext}$.}
    \label{fig:inp_length}
  \end{minipage}
  \hfill
  \begin{minipage}{0.47\textwidth}
    \centering
    \includegraphics[trim={1.75mm 2mm 0 0},clip]{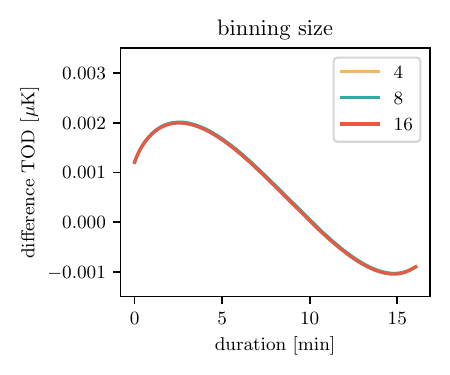}
    \caption{Synthetic extensions for a few different choices of the binning factor, $n_\text{bin}$.}
    \label{fig:bin_factor}
  \end{minipage}
\end{figure}

\subsubsection{Auxillary pixels}\label{subsubsec:poitings}
\begin{itemize}[topsep=2mm,parsep=1mm,itemsep=1mm,leftmargin=6mm]
    \item[$\triangleright$] We append $n_\mathrm{ext}$ samples to the pixel timeline chunk and assign them to a dedicated auxiliary pixel that does not correspond to any physical sky location.
    \item[$\triangleright$] We similarly append $n_\mathrm{ext}$ angles, $\psi$, to the polarization angle timeline chunk. The synthetic angles are uniformly spaced over the interval $[0,\pi)$, ensuring that polarization can be solved for the auxiliary pixels.
\end{itemize}
We repeat this procedure for every detector and TOD chunk. The resulting extended pointing matrix contains \(n_\mathrm{pix}+n_\mathrm{det}n_\mathrm{chunk}\) pixels, of which only the first \(n_\mathrm{pix}\) correspond to physical sky locations.

{This construction is approximate because the synthetic samples contain contributions from the sky signal, which may vary across the extended region. Assigning all synthetic samples in a chunk to a single auxiliary pixel effectively assumes that this contribution can be represented by a single set of Stokes parameters. A possible refinement would be to assign multiple auxiliary pixels to each chunk, allowing a more flexible representation of the sky-signal contribution. However, we find that the resulting bias is much smaller than the edge effects mitigated by STITCH, so we do not pursue this refinement further. 
}

\subsubsection{Implementation in \texttt{BrahMap}}\label{subsubsec:brahmap}
Once the TOD and pointing are extended using the procedure introduced above, the resulting data can be processed with the standard GLS map-making procedure. The additional samples correspond to auxiliary pixels introduced by the synthetic extension; the corresponding pixels are therefore discarded after map-making, leaving only the region of interest in the final map.

We have implemented this procedure in the \texttt{BrahMap} \cite{Anand:2025qio}: a modular map-making framework for CMB experiments that provides the numerical infrastructure for solving the GLS map-making problem. In this work, we integrate STITCH into \texttt{BrahMap}, allowing the synthetically extended timelines and pointing information to be processed using the existing GLS solver. The specific implementation details will be documented alongside the code when it is released.

\subsection{Recovery of the reference GLS solution}\label{subsec:noise}
We test whether STITCH can recover the GLS solution obtained when the circulant approximation is valid. To this end, we consider a LiteBIRD-like scanning strategy (see table~\ref{tab:scan_params} for the specific parameters). We assume a (circular Gaussian) beam full-width half-maximum (FWHM) of $20~\mathrm{arcmin}$, representative of a LiteBIRD CMB channel, and use two pairs of orthogonal detectors. The first pair ($0^\circ$ and $90^\circ$ polarization angles) is located at the telescope boresight, while the second pair ($45^\circ$ and $135^\circ$ polarization angles) is slightly shifted along the focal plane's $x$ axis, which is roughly aligned with the scanning direction. On the sky, the pointings of the two pairs are separated by $\offsetarcmin~\mathrm{arcmin}$, corresponding to 2.5 beam FWHMs. Our choice if using 4 detectors with $0^\circ$, $90^\circ$, $45^\circ$ and $135^\circ$ polarization angles is optimal for map-making. Adding more detectors should result in a global rescaling of the noise angular power spectrum. We simulate one year (\(365.25\) days) of observations. For Fourier-space synthesis of stationary \(1/f\) noise realizations with circulant covariance, we choose the total number of samples per detector to be a power of two, \(n_\text{samp}=2^n\). The smallest value of \(n\) that provides adequate beam sampling\footnote{Neglecting the (slower) precession, the characteristic angular velocity of LiteBIRD's scanning strategy is $\omega_\text{LB} \equiv \sin\beta\,\omega_\text{spin}\simeq90~\mathrm{arcmin}/\mathrm{s}$. It therefore takes LiteBIRD roughly \(t_\text{beam}\equiv20~\mathrm{arcmin}/\omega_\text{LB}\simeq0.2~\mathrm{s}\) to traverse one beam FWHM. A sampling rate \(f_\text{samp}\ge15~\mathrm{Hz}\) provides \(\gtrsim3\) samples per beam-FWHM crossing, which is sufficient to sample the beam adequately.} is \(n=29\), corresponding to a sampling frequency of \(f_\text{samp}\simeq\fsampHz~\mathrm{Hz}\). Adequate sampling of the beam on the sky also informs our choice of the output HEALPix resolution, \(N_\text{side}=512\). 

We simulate the noise-only TOD over 16 segments of approximately 23 days each\footnote{We chose long segments to reduce the influence of the segment boundaries on the reconstructed maps. We also tested shorter segments and found no significant differences in the reconstructed maps for the configurations considered.}. For each segment, we produce a circulant \(1/f\) noise realization with \(f_\text{knee}=100~\mathrm{mHz}\) and \(\slopef=2\). We then construct three sets of maps:
\begin{description}[leftmargin=6mm, itemsep=0pt]
\item[\normalfont$\widehat{\mathbf{m}}_\text{ref}$:] As a reference case, we run the circulant GLS map-maker segment by segment, without the inpainting step. Since the noise realization is circulant over each segment, the assumed covariance matches the simulated one, and the GLS returns the optimal solution.
\item[\normalfont$\widehat{\mathbf{m}}_\text{chunks}$:] We divide each TOD circulant segment into 16 shorter chunks, each $\sim 1.2$ day long\footnote{We choose the chunk boundaries so that the sky positions observed at the two ends of each chunk are well separated. This reduces the risk that the periodic boundary condition assumed by the circulant noise covariance introduces artificial coupling between nearby sky regions.} and process them with the circulant GLS map-maker. We expect large-scale noise residuals, since the circulant approximation is no longer valid for the shorter chunks, leading to the edge effects discussed in section~\ref{subsec:edge}.
\item[\normalfont$\widehat{\mathbf{m}}_\text{STITCH}$:] In this case, instead of using the circulant GLS map-maker right away, we first preprocess the TOD chunks with STITCH using \(n_\mathrm{ext}=2^{14}\), corresponding to roughly 16 minutes of data, and \(n_\text{bin}=8\). We expect this procedure to recover maps close to the reference solution, \(\widehat{\mathbf{m}}_\text{ref}\).
\end{description}
{Figure~\ref{fig:nostat} shows the Stokes $Q$ maps obtained with the three map-making approaches, qualitatively illustrating how applying a circulant map-maker to non-circulant chunks produces large-scale residuals that are mitigated by STITCH preprocessing. We quantify these residuals further below.}

\begin{figure}[!t]
    \centering
    \includegraphics{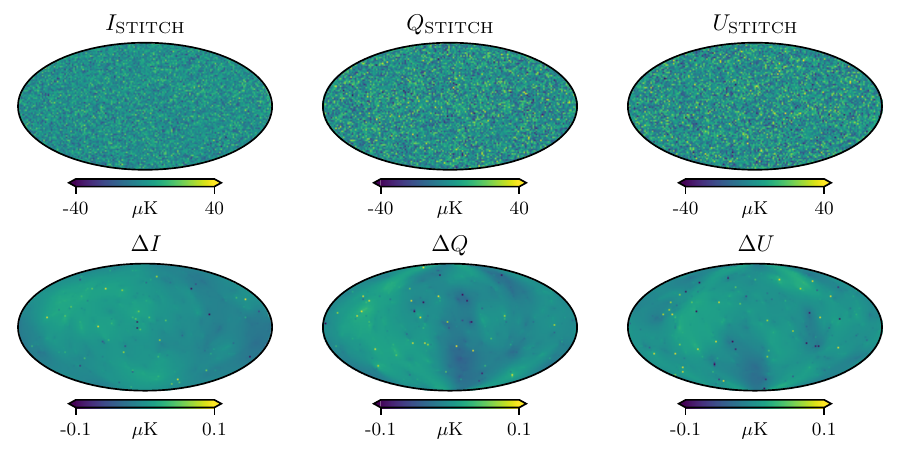}
    \caption{Impact of STITCH on the noise maps. The top row shows the $I$, $Q$, and $U$ maps obtained by preprocessing a noise-only TOD with STITCH ($f_{\rm knee}=100\,{\rm mHz}$ and $\slopef = 2$) and making maps with a circulant GLS. The bottom row shows the corresponding residual $\Delta \mathbf{m} \equiv \widehat{\mathbf{m}}_\text{STITCH}-\widehat{\mathbf{m}}_\text{ref}$. Note the different ranges spanned by the color bars.}
    \label{fig:diff_maps}
\end{figure}

\begin{figure}
    \centering
    \includegraphics[trim={1.5mm 5mm 12.5mm 15mm},clip]{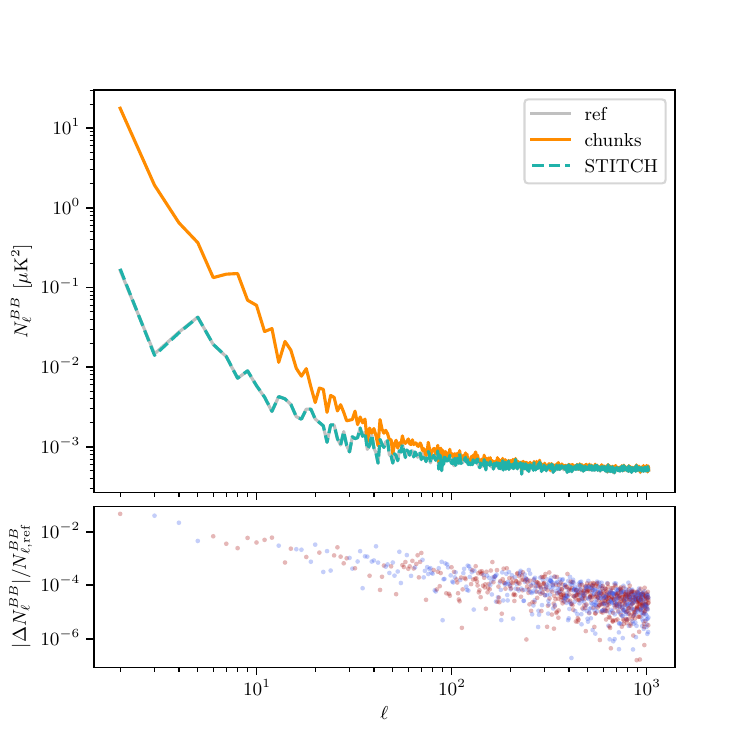}
    \caption{Top panel: angular $BB$ power spectra obtained from the same $1/f$ noise realization, which is circulant over $\sim 23$ day segments. The different lines were obtained with different map-making approaches. The reference solution (gray) uses a circulant GLS over the fulls circulant segments. The other two lines are instead obtained by processing the noise over $\sim 1.2$ day chunks with and without STITCH preprocessing (dashed teal and orange, respectively). The breaking of the circulant assumption introduced by processing the data in chunks leads to an excess of large-scale residual noise, visible as a low-$\ell$ excess in the orange curves. Preprocessing with STITCH recovers a residual spectrum consistent with the reference solution. Bottom panel: absolute value of the relative difference between the STITCH and reference spectra: red for positive differences and blue for negative.
}
    \label{fig:cls_together}
\end{figure}

The \(I\), \(Q\), and \(U\) maps obtained with STITCH are shown in figure~\ref{fig:diff_maps}, together with the corresponding difference maps, \(\Delta\widehat{\mathbf{m}}\equiv\widehat{\mathbf{m}}_\text{STITCH}-\widehat{\mathbf{m}}_\text{ref}\), while figure~\ref{fig:cls_together} compares the \(BB\) angular noise power spectra obtained with the three map-makers. Relative to the reference reconstruction (gray line), processing the circulant segments in chunks (orange line) produces excess large-scale noise because of edge effects. Preprocessing the chunks with STITCH before applying the circulant GLS (dashed teal line) suppresses these residuals, yielding a spectrum that closely matches the reference result.
The good agreement between the reference and STITCH spectra also shows that neglecting information at temporal scales between $\sim 1.2$ and $\sim 23$ days have very small impact on the recovered solution.

\subsection{Impact on CMB reconstruction}\label{subsec:CMB}
Having shown that STITCH suppresses the edge effects introduced by processing data in chunks, we now assess whether the inpainting procedure introduces biases in the reconstructed CMB signal. Because both STITCH preprocessing and GLS map-making are linear operations, the noise and CMB contributions can be considered separately. We therefore repeat the analysis of section~\ref{subsec:noise} using noise-free CMB simulations. We use a lensed CMB realization with no primordial $B$ modes, $r=0$, in input.

For this validation, we assumed the sky scanning centered at each pixel. This removes sub-pixel effects and allows us to isolate the impact of the inpainting procedure; the extension to realistic detector pointings is discussed in section~\ref{subsec:subpix}. As in section~\ref{subsec:noise}, we consider the three map-making cases ($\widehat{\mathbf{m}}_\text{ref}$, $\widehat{\mathbf{m}}_\text{chunks}$ and $\widehat{\mathbf{m}}_\text{STITCH}$). The STITCH preprocessing is done assuming $f_\text{knee} = 100~\mathrm{mHz}$ and $\slopef=2$. We use the difference map
\begin{equation}\label{eqn:deltastitch}
    \Delta\mathbf{m} = \widehat{\mathbf{m}}_\text{STITCH} - \widehat{\mathbf{m}}_\text{ref}\,,
\end{equation}
to quantify the residual due to the preprocessing. The GLS itself (whether it is applied to full segments or chunks) does not bias the sky signal, so any residual in $\Delta\mathbf{m}$ is due to signal leaking in the synthetic data.

\begin{figure}[!t]
    \centering
    \includegraphics{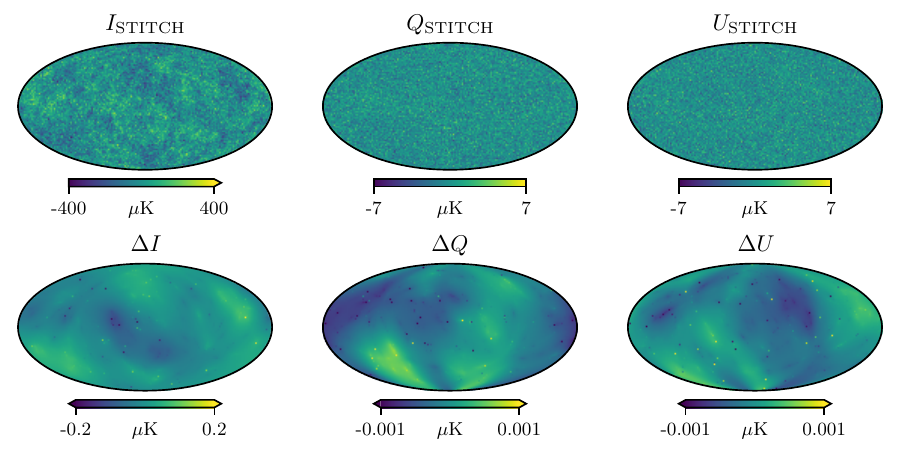}
    \caption{Impact of STITCH on the reconstructed CMB. The top row shows the $I$, $Q$, and $U$ maps obtained by observing a CMB-only sky, preprocessing the TOD with STITCH ($f_{\rm knee}=100\,{\rm mHz}$ and $\slopef = 2$) and making maps with a circulant GLS. The bottom row shows the corresponding residual $\Delta \mathbf{m} \equiv \widehat{\mathbf{m}}_\text{STITCH}-\widehat{\mathbf{m}}_\text{ref}$. Notice the different ranges spanned by the color bars.}
    \label{fig:res_NOinp_inp_CMB}
\end{figure}

Figure ~\ref{fig:res_NOinp_inp_CMB} shows the residual maps. The inpainting procedure introduces visible large-scale features, of $\mathcal{O}(10^{-4})$ level as compared to the input, and point-like residuals near the chunk boundaries. However, their amplitudes are much smaller than those of the input CMB signal, as indicated by the different color scales. These residuals come from $E\to B$ leakage, as $I\to P$ leakage is zero by construction because of the choice of using perfectly orthogonal detector pairs.

\begin{figure}[!t]
    \centering
    \includegraphics{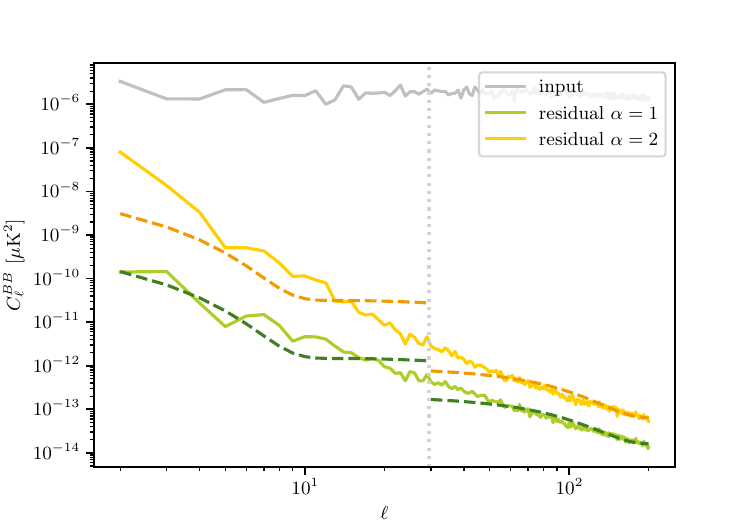}
    \caption{Angular $BB$ power spectra of the residual CMB maps, $\Delta \mathbf{m} \equiv \widehat{\mathbf{m}}_\text{STITCH}-\widehat{\mathbf{m}}_\text{ref}$, for $\slopef=2$ (yellow) and $\slopef=1$ (light green). In both cases, $f_\text{knee}=100~\mathrm{mHz}$. The dashed lines show the corresponding best-fit curves over the low- and high-multipole ranges. For reference, the angular power spectrum of the input CMB is shown in gray.}
    \label{fig:res_spectra_CMB}
\end{figure}
 
The impact of these residuals is more naturally quantified in harmonic space. We therefore compute their $BB$ angular power spectrum and fit it for $r$ and $A_\text{lens}$ over the two multipole ranges $2\leq\ell\leq29$ and $30\leq\ell\leq200$. These fits provide an effective measure of the potential impact of the residuals on primordial and lensing $B$-mode measurements. We obtain $\{r=2.0\times10^{-7},\,A_\text{lens}=0.0\}$, over the low-multipole range, and $\{r=5.2\times10^{-9},\,A_\text{lens}=2.4\times10^{-8}\}$, over the high-multipole range. The residual spectrum and the corresponding best-fit models are shown by the solid yellow and dashed orange curves in figure~\ref{fig:res_spectra_CMB}, respectively.
To assess the dependence on the assumed noise model, we repeat the analysis using a less aggressive $1/f$ spectrum with $\slopef=1$. As shown by the light-green curve, the residual $BB$ spectrum is approximately two orders of magnitude smaller than for $\slopef=2$.

The inpainting procedure introduces a negligible residual in the reconstructed CMB signal, while substantially reducing the large-scale noise induced by the edge-effects.

\section{Mapping temporal $1/f$ noise onto the sky}\label{sec:scanstr}
Having validated that STITCH preprocessing enables standard circulant GLS solvers to recover the reference noise solution without introducing significant bias into the reconstructed CMB, we now apply this pipeline to investigate how temporal $1/f$ noise projects onto angular sky scales across different satellite scanning strategies. Specifically, we examine the empirical relation between the temporal noise spectrum $P(f)$ and the effective angular noise spectrum $N_\ell^{\rm BB}$. 
The relation between these two quantities depends not only on the temporal noise properties, but also on the satellite scanning strategy, which determines how temporal frequencies contribute to different angular scales.

\begin{figure}[!t]
    \centering
    \small
    \begin{tikzpicture}
        \draw[gray!50,-latex] (-10,0) -- (2.5,0);
        \draw[fill=yellow!70!pink] (-10,0) circle (0.5);
        \node at (-10,0.8) {Sun};
        \draw[fill=teal!40!green!80!blue!80!cyan] (-3,0) circle (0.3);
        \node at (-3,0.6) {Earth};
        \begin{scope} [rotate=45]
            \fill[yellow!50!orange, opacity=0.4] (0,0) -- (0.75,0) arc (360:315:0.75) -- (0,0);
            \draw (0.75,0) arc (360:315:0.75);
            \draw[-latex] (0,0) -- (2.5,0);
            \begin{scope} [shift={(0.4,0)},rotate=50]
                \fill[red!50!orange, opacity=0.4] (0,0) -- (0.5,0) arc (360:310:0.5) -- (0,0);
                \draw (0.5,0) arc (360:310:0.5);
                \draw[-latex] (0,0) -- (2.25,0);
                \draw[fill=gray!80!white] (-.1,-.07) rectangle (.25,.07);
            \end{scope}
            \draw[fill=gray] (-.4,-.2) rectangle (.4,.2);
            \draw[-latex] (2,.2) arc (20:340:.25 and .5);
        \end{scope}
        \draw[-latex] (2,.2) arc (20:340:.25 and .5);
        \node at (1.1,0.35) {$\alphascan$};
        \node at (0.55,1.1) {$\betascan$};
        \node[text width=2cm,align=flush left] at (2.8,1) {$\omega_\text{spin}$};
        \node[text width=2cm,align=flush left] at (3.1,-.5) {$\omega_\text{prec}$};
        \node at (.2,-.8) {spacecraft};
    \end{tikzpicture}
    \caption{Intuitive definition of a CMB spacecraft's scanning parameters. The anti-Sun direction (light gray arrow) forms an angle $\alphascan$ with the spin axis which, in turn, forms an angle $\betascan$ with the boresight. The spacecraft precesses while spinning, with angular frequencies $\omega_\text{prec}$ and $\omega_\text{spin}$, respectively.}
    \label{fig:LITEsketch}
\end{figure}
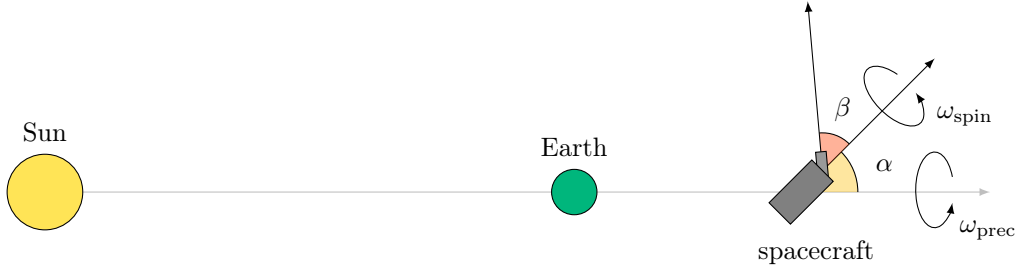

\begin{figure}[!t]
    \centering
    \includegraphics{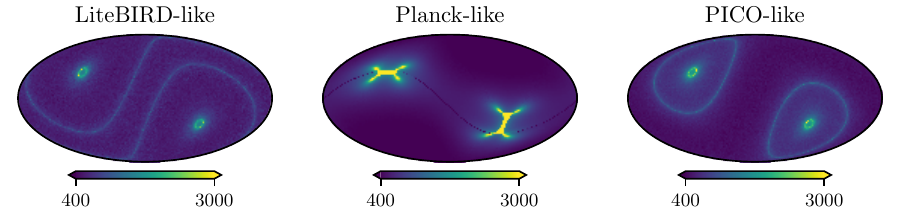}
    \caption{Hitmaps in Galactic coordinates for the three representative scanning strategies specified in table~\ref{tab:scan_params}. In all cases, the sky is scanned by two detector pairs (one at boresight and offset of $\sim \offsetarcmin$ arcmin along the scanning direction) with $f_\text{samp}\simeq\fsampHz~\mathrm{Hz}$, for 1 year.}
    \label{fig:hitmap}
\end{figure}

We characterize this temporal-to-angular mapping for a range of temporal noise parameters, \(f_\text{knee}\in \{5,10,25,50,100\}\,\mathrm{mHz}\) and \(\slopef\in\{1,2\}\), as introduced in figure~\ref{fig:noise_Pf}, and for three representative spinning-satellite scanning strategies. For a spinning CMB satellite, the scanning strategy is determined by the spin rate, the precession rate, and the orientation of the spin axis with respect to the boresight and anti-Sun direction, as illustrated in figure~\ref{fig:LITEsketch}. We consider configurations representative of \emph{Planck}, LiteBIRD, and PICO. The corresponding scanning parameters are summarized in table~\ref{tab:scan_params}, together with the maximum angular velocity, \(\omega_{\rm max}\), defined as \cite{LiteBIRD:2024fzn}
\begin{table}[!t]
    \centering
    \begin{tabular}{r|cccc|c}
         & $\alphascan$ & $\betascan$ & $T_\text{prec}$ & $f_\text{spin}\,[{\rm rpm}]$ & $\omega_{\rm max}\,[{}^\circ/{\rm s}]$\\
        \hline
        LiteBIRD-like & $37.5^\circ$ & $57.5^\circ$ & 192.348 min & 0.3 & 1.6\\
        \emph{Planck}-like & $7.5^\circ$ & $85.0^\circ$ & 6 months & 1.0 & 6.1\\
        PICO-like & $26.0^\circ$ & $69.0^\circ$ & 10 hours & 1.0  & 5.6
    \end{tabular}
    \caption{Scanning strategy parameters for the three representative setups we consider in this paper \cite{LiteBIRD:2022cnt, LiteBIRD:2026gdf, Dupac:2004pg, NASAPICO:2019thw}. The angles $\alphascan$ and $\betascan$ as illustrated in figure~\ref{fig:LITEsketch}, together with the corresponding value of $\omega_{\rm max}$ defined as in eq.~\eqref{eqn:omegamax}.}
    \label{tab:scan_params}
\end{table}
\begin{equation}\label{eqn:omegamax}
\omega_{\rm max} = \omega_\alphascan\sin(\alphascan + \betascan) + \omega_\betascan\sin\betascan\,,
\end{equation}
which we use to characterize the scan characteristic angular velocity. The differences in the scanning strategies of these three missions are illustrated by their corresponding hitmaps, shown in figure~\ref{fig:hitmap}.

For each combination of temporal noise parameters and scanning strategy, we generate 25 independent noise-only simulations and compute the resulting \(BB\) angular noise spectra. We then fit these spectra with an effective angular noise model and use the best-fit parameters to derive empirical relations between the temporal and angular descriptions of the noise. These relations provide a practical way to translate detector-level \(1/f\) noise properties into their expected impact on the sky and to compare the noise performance of different scanning strategies.

We adopt the same instrumental configuration as in section~\ref{subsec:noise}, except that both detector pairs are placed at boresight. We expect a shift along the focal-plane \(x\) axis not to affect the projected noise, as the scanning strategy is nearly equivalent for the two pairs, and we have verified that this is the case.

%

\subsection{Fitting the angular noise models}\label{subsec:best_fit}
We characterize the simulated angular noise spectra using an effective model that mirrors the functional form adopted for the temporal noise in eq.~\eqref{eqn:Pf},
\begin{equation}\label{eqn:noise_spectra_ell}
    N_\ell^{BB}(\wnl, \ell_\text{knee},\slopel) = \wnl \left[1+\left(\frac{\ell_\text{knee}}{\ell}\right)^{\slopel}\right],
\end{equation}
where $\wnl$ parametrizes the white-noise level, while $\ell_\mathrm{knee}$ and $\slopel$ play roles analogous to $f_\mathrm{knee}$ and $\slopef$ in temporal domain. These parameters provide an effective way to summarize the angular noise spectrum and, in particular, to quantify how the temporal noise parameters are projected onto the sky.

We determine the best-fit parameters by maximizing the likelihood of the simulated $BB$ spectra.
Under the general assumption of statistically isotropic Gaussian fluctuations, the probability of observing an angular power spectrum realization $\widehat{C}_{\ell}$ given an underlying spectrum $C_\ell$ is \cite{Hamimeche:2008ai}
\begin{equation}\label{eqn:logPchp3}
\log P_{\ell}
\left(
\widehat{C}_{\ell}
\,\big|\,
C_\ell
\right)
=
-\frac{2\ell+1}{2}
\left[
\log C_\ell
+
\frac{\widehat{C}_{\ell}}{C_\ell}
-
\frac{2\ell-1}{2\ell+1}
\log\widehat{C}_{\ell}
\right]
+\mathrm{const}.\,.
\end{equation}
If $C_\ell$ is parameterized by a set of parameters $(p_1,p_2,\ldots)$, and the estimated power spectra at different multipoles are treated as statistically independent, eq.~\eqref{eqn:logPchp3} gives the following likelihood:
\begin{equation}\label{eqn:logL}
\log L(p_1,p_2,\ldots) \equiv 
-\sum_{\ell}\frac{2\ell+1}{2}
\left[
\log C_\ell
+
\frac{\widehat{C}_{\ell}}{C_\ell}
\right]
+\mathrm{const}.\,.
\end{equation}
By maximizing the likelihood against $\{p_1,p_2,\ldots\}$, one can extract the best-fit parameters. In our case, however, for each combination of scanning strategy and set of parameters $\{\sigma, f_\mathrm{knee}, \slopef\}$, we simulate a set of 25 independent noise angular power spectra, $\widehat{N}_{\ell,i}^{BB}$, spanned by the index $i$. Being independent, they all contribute the same to the overall likelihood and we can write
\begin{equation}\label{eqn:logLi}
\log L(p_1,p_2,\ldots) \equiv
-\sum_{\ell}\frac{2\ell+1}{2}
\left[
\log N_\ell^{BB}
+
\frac{\widehat{N}_{\ell,\text{avg}}^{BB}}{N_\ell^{BB}}
\right]
+\mathrm{const}.\,,
\end{equation}
where $\widehat{N}_{\ell,\text{avg}}^{BB}\equiv\sum_i\widehat{N}_{\ell,i}^{BB}/25$. 

\begin{figure}
    \centering
    \includegraphics[trim={3mm 4mm 0 3mm},clip]{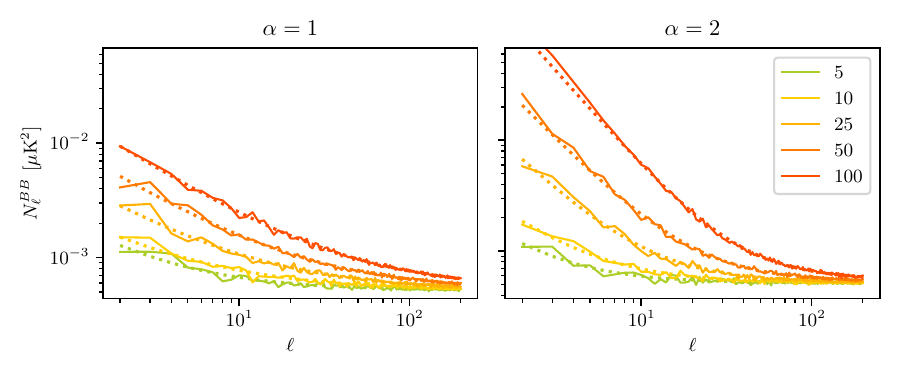}
    \caption{Noise $BB$ angular power spectrum for a LiteBIRD-like scanning strategy. The solid lines show the simulated $\widehat{N}_\ell^{BB}$ averaged over 25 independent noise realizations for different choices of noise parameters. The corresponding best-fit curves given the model in eq.~\eqref{eqn:noise_spectra_ell} are shown in dotted.}
    \label{fig:bestfitNl}
\end{figure}

The model in eq.~\eqref{eqn:noise_spectra_ell} provides a good description of the simulated spectra. Figure~\ref{fig:bestfitNl} compares the average $BB$ noise spectra for a LiteBIRD-like scanning strategy with their corresponding best-fit curves. Although eq.~\eqref{eqn:noise_spectra_ell} is intended as an effective description of the large-scale correlated-noise excess rather than an exact representation of the simulated spectra, it captures their scale dependence well over the multipole range of interest. This motivates using $\ell_\mathrm{knee}$ and $\slopel$ as effective angular counterparts of the temporal noise parameters $f_\mathrm{knee}$ and $\slopef$.

The best-fit values of $\ell_\mathrm{knee}$ and $\slopel$ for all combinations of temporal noise parameters and scanning strategies considered in this work are reported in table~\ref{tab:lknee}. The quoted uncertainties are obtained by fitting eq.~\eqref{eqn:noise_spectra_ell} independently to each of the 25 simulated realizations and taking the realization-to-realization scatter of the resulting best-fit parameters. For compactness, we do not report the best-fit values of $\wnl$ in the table. We also find no significant variation in the fitted parameters when the noise maps are rescaled by the square root of the number of hits per pixel, suggesting that our results are insensitive to this normalization.

The table already reveals two systematic trends: increasing $f_\mathrm{knee}$ shifts $\ell_\mathrm{knee}$ to higher multipoles, while increasing $\slopef$ generally produces a steeper angular spectrum. The magnitude of both effects also depends on the scanning strategy. We comment on these dependencies separately below.

\begin{table}[!t]
\centering
\begin{tabular}{|c|c|c|c|c|c|c|}
\hline
$f_\mathrm{knee}$
& \multicolumn{2}{c|}{LiteBIRD-like}
& \multicolumn{2}{c|}{\emph{Planck}-like}
& \multicolumn{2}{c|}{PICO-like} \\
$[\mathrm{mHz}]$ & $\slopef=1$ & $\slopef=2$
& $\slopef=1$ & $\slopef=2$
& $\slopef=1$ & $\slopef=2$ \\
\hline
\multirow{2}{*}{5} & \cellcolor{RedOrange!27.3!Goldenrod}$2.9\pm0.2$ & \cellcolor{RedOrange!22.2!Goldenrod}$2.4\pm0.2$ & \cellcolor{RedOrange!29.6!Goldenrod}$3.1\pm0.2$ & \cellcolor{RedOrange!25.0!Goldenrod}$2.7\pm0.2$ & \cellcolor{RedOrange!0.0!Goldenrod}$1.1\pm0.2$ & \cellcolor{RedOrange!7.3!Goldenrod}$1.4\pm0.2$ \\
  & \cellcolor{RedOrange!27.3!Goldenrod}$1.0\pm0.7$ & \cellcolor{RedOrange!22.2!Goldenrod}$1.3\pm0.1$ & \cellcolor{RedOrange!29.6!Goldenrod}$1.5\pm0.6$ & \cellcolor{RedOrange!25.0!Goldenrod}$1.7\pm0.7$ & \cellcolor{RedOrange!0.0!Goldenrod}$1.0\pm0.8$ & \cellcolor{RedOrange!7.3!Goldenrod}$2\pm1$ \\
\hline
\multirow{2}{*}{10} & \cellcolor{RedOrange!39.1!Goldenrod}$4.4\pm0.3$ & \cellcolor{RedOrange!37.9!Goldenrod}$4.2\pm0.3$ & \cellcolor{RedOrange!34.3!Goldenrod}$3.7\pm0.3$ & \cellcolor{RedOrange!31.1!Goldenrod}$3.3\pm0.2$ & \cellcolor{RedOrange!13.6!Goldenrod}$1.8\pm0.3$ & \cellcolor{RedOrange!7.1!Goldenrod}$1.4\pm0.2$ \\
  & \cellcolor{RedOrange!39.1!Goldenrod}$0.83\pm0.05$ & \cellcolor{RedOrange!37.9!Goldenrod}$1.27\pm0.07$ & \cellcolor{RedOrange!34.3!Goldenrod}$1.2\pm0.1$ & \cellcolor{RedOrange!31.1!Goldenrod}$2.1\pm0.4$ & \cellcolor{RedOrange!13.6!Goldenrod}$1.0\pm0.2$ & \cellcolor{RedOrange!7.1!Goldenrod}$1.4\pm0.9$ \\
\hline
\multirow{2}{*}{25} & \cellcolor{RedOrange!63.9!Goldenrod}$10.8\pm0.4$ & \cellcolor{RedOrange!63.7!Goldenrod}$10.7\pm0.2$ & \cellcolor{RedOrange!45.2!Goldenrod}$5.5\pm0.3$ & \cellcolor{RedOrange!40.9!Goldenrod}$4.7\pm0.3$ & \cellcolor{RedOrange!29.9!Goldenrod}$3.2\pm0.2$ & \cellcolor{RedOrange!30.3!Goldenrod}$3.2\pm0.2$ \\
  & \cellcolor{RedOrange!63.9!Goldenrod}$0.88\pm0.03$ & \cellcolor{RedOrange!63.7!Goldenrod}$1.48\pm0.02$ & \cellcolor{RedOrange!45.2!Goldenrod}$1.17\pm0.06$ & \cellcolor{RedOrange!40.9!Goldenrod}$1.5\pm0.1$ & \cellcolor{RedOrange!29.9!Goldenrod}$0.91\pm0.05$ & \cellcolor{RedOrange!30.3!Goldenrod}$1.25\pm0.08$ \\
\hline
\multirow{2}{*}{50} & \cellcolor{RedOrange!82.6!Goldenrod}$21.2\pm0.7$ & \cellcolor{RedOrange!81.3!Goldenrod}$20.3\pm0.2$ & \cellcolor{RedOrange!57.8!Goldenrod}$8.7\pm0.3$ & \cellcolor{RedOrange!60.2!Goldenrod}$9.5\pm0.3$ & \cellcolor{RedOrange!50.7!Goldenrod}$6.7\pm0.3$ & \cellcolor{RedOrange!48.2!Goldenrod}$6.2\pm0.2$ \\
  & \cellcolor{RedOrange!82.6!Goldenrod}$0.92\pm0.02$ & \cellcolor{RedOrange!81.3!Goldenrod}$1.57\pm0.02$ & \cellcolor{RedOrange!57.8!Goldenrod}$1.02\pm0.04$ & \cellcolor{RedOrange!60.2!Goldenrod}$1.50\pm0.05$ & \cellcolor{RedOrange!50.7!Goldenrod}$0.90\pm0.03$ & \cellcolor{RedOrange!48.2!Goldenrod}$1.41\pm0.07$ \\
\hline
\multirow{2}{*}{100} & \cellcolor{RedOrange!100.0!Goldenrod}$40\pm1$ & \cellcolor{RedOrange!98.5!Goldenrod}$37.6\pm0.3$ & \cellcolor{RedOrange!73.8!Goldenrod}$15.5\pm0.5$ & \cellcolor{RedOrange!81.3!Goldenrod}$20.3\pm0.4$ & \cellcolor{RedOrange!67.9!Goldenrod}$12.5\pm0.4$ & \cellcolor{RedOrange!67.5!Goldenrod}$12.3\pm0.2$ \\
  & \cellcolor{RedOrange!100.0!Goldenrod}$0.93\pm0.01$ & \cellcolor{RedOrange!98.5!Goldenrod}$1.74\pm0.01$ & \cellcolor{RedOrange!73.8!Goldenrod}$0.97\pm0.02$ & \cellcolor{RedOrange!81.3!Goldenrod}$1.49\pm0.03$ & \cellcolor{RedOrange!67.9!Goldenrod}$0.90\pm0.03$ & \cellcolor{RedOrange!67.5!Goldenrod}$1.59\pm0.03$ \\
\hline
\end{tabular}
\caption{Best-fit values of the $\ell_\mathrm{knee}$ and $\slopel$ parameters for different choices of noise temporal parameters and scanning strategies. Each cell shows two numbers: $\ell_\mathrm{knee}$ on top and $\slopel$ at the bottom. The color of the cell depends on $\ell_\mathrm{knee}$: yellow for lower values and red for higher.}
\label{tab:lknee}
\end{table}

\begin{figure}[!t]
    \centering
    \includegraphics[trim={3mm 4mm 0 3mm},clip]{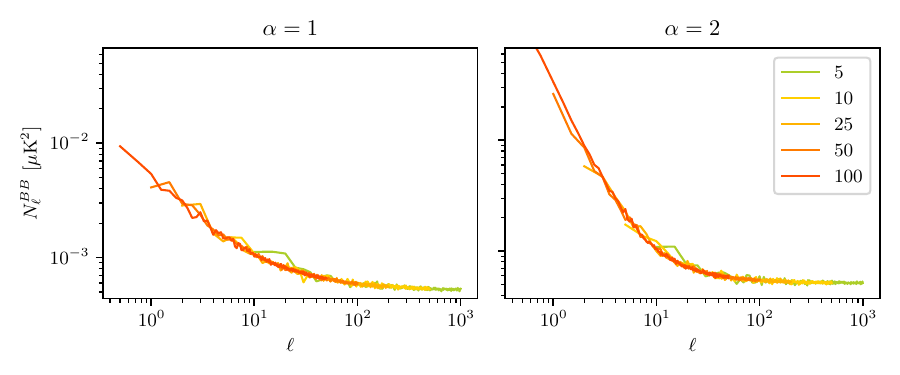}
    \caption{Same average noise $BB$ angular power spectra for a LiteBIRD-like scannning strategy as in figure~\ref{fig:bestfitNl}, but plotted against the rescaled multipole $\ell/(f_{\rm knee}/25\,\mathrm{mHz})$. This rescaling accounts for the approximately linear dependence of the angular knee scale on the temporal knee frequency, allowing the spectra with different $f_{\rm knee}$ to be compared on a common effective multipole scale.}
    \label{fig:Nl_rescaled}
\end{figure}

\paragraph{Dependence on $f_{\rm knee}$}
For a fixed scanning strategy and temporal slope, increasing $f_{\rm knee}$ shifts the onset of correlated noise towards higher multipoles. This is expected because a higher temporal knee frequency extends the correlated-noise contribution to higher temporal frequencies, which are projected onto smaller angular scales by the scan. The fitted values in table~\ref{tab:lknee} show that $\ell_\mathrm{knee}$ is approximately proportional to $f_\mathrm{knee}$ over the range considered here. This scaling is also apparent when the spectra are plotted against the rescaled multipole $\ell/(f_{\rm knee}/25\,\mathrm{mHz})$, as shown in figure~\ref{fig:Nl_rescaled}. The spectra for different $f_{\rm knee}$ values approximately overlap after this rescaling, confirming that the dependence of the angular knee on $f_{\rm knee}$ is close to linear.

Motivated by the approximately linear scaling of $\ell_{\rm knee}$ with $f_{\rm knee}$, we combine the spectra at different $f_{\rm knee}$ values into a single fit. Specifically, we use the model
\begin{equation}\label{eqn:noise_spectra_ell_together}
    N_\ell^{BB}(W,\ell_{25},\slopel) = W \left[1+\left(\frac{\ell_{25}}{\ell}\frac{f_\text{knee}}{25\,\mathrm{mHz}}\right)^{\slopel}\right].
\end{equation}
where $W$, $\ell_{25}$, and $\slopel$ are common to all spectra for a given scanning strategy and temporal slope. Here, $\ell_{25}$ represents the effective angular knee at the reference frequency $f_{\rm knee}=25\,\mathrm{mHz}$. We determine these three parameters by jointly maximizing the likelihood over the average $BB$ spectra for all five values of $f_\mathrm{knee}$. The resulting best-fit parameters are reported in table~\ref{tab:best-fit_together}. \minor{Analogously to the case of table~\ref{tab:lknee}, the uncertainties are obtained by fitting eq.~\eqref{eqn:noise_spectra_ell_together} independently to each of the 25 simulated realizations and evaluating the dispersion of the best-fit parameters.}

\begin{table}[!t]
    \centering
    \begin{tabular}{cc|ccc}
       & $\alpha$ & $W\,[\times 10^{-4}]$ & $\ell_{25}$ & $\beta$ \\
       \hline
      \multirow{2}{*}{LiteBIRD} & $1$ & $5.177\pm0.007$ & $10.89\pm0.07$ & $0.888\pm0.005$ \\
                                & $2$ & $5.235\pm0.005$ & $10.76\pm0.05$ & $1.527\pm0.008$ \\
      \hline
      \multirow{2}{*}{\emph{Planck}} & $1$ & $6.375\pm0.008$ & $4.26\pm0.07$ & $0.942\pm0.012$ \\
                                     & $2$ & $6.389\pm0.004$ & $5.09\pm0.07$ & $1.447\pm0.014$ \\
      \hline
      \multirow{2}{*}{PICO} & $1$ & $5.359\pm0.008$ & $3.22\pm0.05$ & $0.869\pm0.011$ \\
                            & $2$ & $5.398\pm0.005$ & $3.24\pm0.03$ & $1.406\pm0.021$ \\
    \end{tabular}
    \caption{Best-fit parameters of the collective angular noise model, eq.~\eqref{eqn:noise_spectra_ell_together}, obtained by jointly fitting the average $BB$ noise spectra for the five values of $f_{\rm knee}$. The fits are performed separately for each scanning strategy and temporal noise slope $\slopef$. \minor{The uncertainties are obtained by fitting each realization independently and taking the scatter of the best-fit parameters.}}
    \label{tab:best-fit_together}
\end{table}

\paragraph{Dependence on $\alpha$}
Steeper temporal noise spectra produce steeper large-scale angular noise spectra. This is reflected in the fitted values of $\slopel$, which increase when the temporal slope changes from $\slopef=1$ to $\slopef=2$. With only two values of $\slopef$ considered here, however, we do not have sufficient sampling to derive a reliable empirical relation between $\slopef$ and $\slopel$. We leave a quantitative empirical relation for future work with a broader range of temporal slopes. Still, we generally observe $\beta<\alpha$. This is a consequence of GLS map-making mitigation.

\paragraph{Dependence on scanning strategy}
The fitted angular parameters summarized in table~\ref{tab:best-fit_together} vary across the three scanning strategies, demonstrating that the mapping from temporal to angular scales is not universal. For a given temporal frequency, faster scanning shifts the corresponding noise contribution towards lower multipoles, approximately following \(\ell\propto f/\omega_{\rm scan}\). {If the scanning speed were the only relevant factor, we would expect the characteristic multipole $\ell_{25}$ to scale inversely with the scanning frequency. Intuitively, observing the same pixel from different scanning directions can provide additional constraints on large-scale correlated noise, potentially improving its suppression during map-making. Thus, the effective parameters $\ell_{25}$ and $\slopel$ would depend on both the characteristic scanning frequency and scan geometry.
To test this expectation, we use $\omega_\text{max}$ defined in eq.~\eqref{eqn:omegamax} as proxy for the characteristic scanning frequency and rescale the best-fit values of $\ell_{25}$ to the LiteBIRD reference value, $\omega_\text{max,LB} = 1.6^\circ/\mathrm{s}$, reported in table~\ref{tab:scan_params}:
\begin{equation}
    \ell_{25,\text{eff}} \equiv \ell_{25} \frac{\omega_\text{max}}{\omega_\text{max,LB}}.
\end{equation}
\minor{For the $\alpha = 1$ case, we find $\ell_{25,\text{eff}} = 16.2\pm0.3$ for \emph{Planck} and $\ell_{25,\text{eff}} = 11.3\pm0.2$ for PICO, compared with the LiteBIRD reference value of $10.89\pm 0.07$ reported in table~\ref{tab:best-fit_together}.} This difference confirms that scan geometry also influences the temporal-to-angular noise mapping. 
In particular, the similar values of $\ell_{25,\mathrm{eff}}$ for PICO and LiteBIRD reflect their more similar scan geometries compared with Planck. This interpretation is consistent with \cite{LiteBIRD:2024fzn}, which reported modest variations in hit-count variance and cross-linking factor across a broad range of scan-geometry parameters, including those considered for our LiteBIRD- and PICO-like configurations.
}

\section{Validation and robustness tests}\label{sec:validation}
In section~\ref{sec:stitch}, we showed how combining STITCH with circulant GLS mitigates the edge effects introduced by processing the TOD in chunks, while introducing negligible residuals into the recovered CMB maps. We now further assess the robustness of the method by examining its impact on foreground reconstruction
and its interplay with sub-pixel effects.

\subsection{Impact of foregrounds}\label{subsec:fgs}
Here we repeat the exercise of section~\ref{subsec:CMB} for Galactic foreground emission. Since CMB and foreground emission have different angular structures, they contribute to different temporal frequencies as the satellite scans the sky. We therefore verify that STITCH does not introduce significant residuals or distortions in the reconstructed foreground maps.

\begin{figure}[!t]
    \centering
    \includegraphics{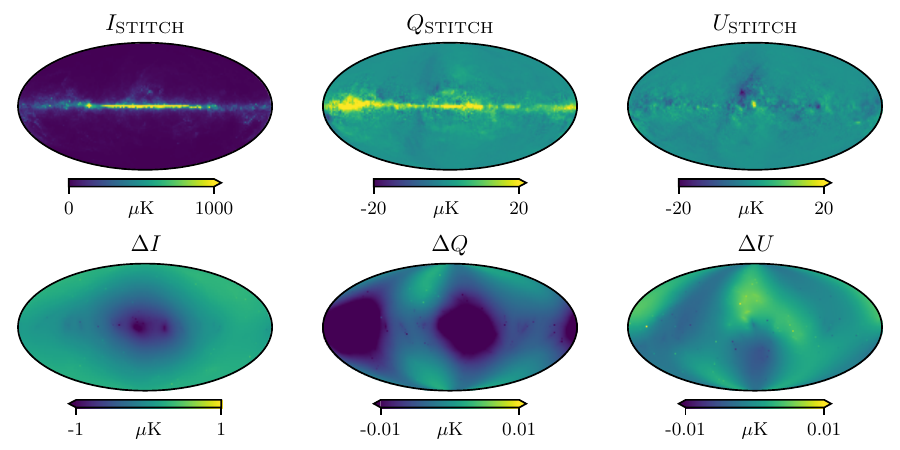}
    \caption{Impact of STITCH on the reconstructed foreground maps. The top row shows the $I$, $Q$, and $U$ maps obtained by observing a \texttt{d0s0} sky, preprocessing the TOD with STITCH ($f_{\rm knee}=100\,{\rm mHz}$ and $\slopef = 2$) and making maps with a circulant GLS. The bottom row shows the corresponding residual $\Delta \mathbf{m} \equiv \widehat{\mathbf{m}}_\text{STITCH}-\widehat{\mathbf{m}}_\text{ref}$. Notice the different ranges spanned by the color bars.}
    \label{fig:res_NOinp_inp_d0s0}
\end{figure}

\begin{figure}[!t]
    \centering
    \includegraphics{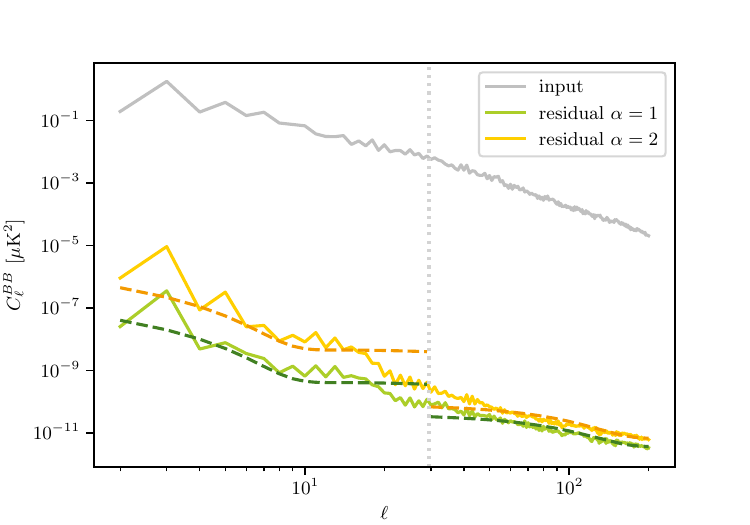}
    \caption{$BB$ angular power spectra of the residual foreground maps, $\Delta \mathbf{m} \equiv \widehat{\mathbf{m}}_\text{STITCH}-\widehat{\mathbf{m}}_\text{ref}$, for $\slopef=2$ (yellow) and $\slopef=1$ (light green). In both cases, $f_\text{knee}=100~\mathrm{mHz}$. The dashed lines show the corresponding best-fit curves over the low- and high-multipole ranges. For reference, the angular power spectrum of the input sky is shown in gray.}
    \label{fig:res_spectra_d0}
\end{figure}

We use the \texttt{PySM}~\cite{Thorne:2016ifb} \texttt{d0s0} sky model as input, including Galactic dust and synchrotron emission. We assume to be observing with a characteristic CMB frequency channel with tophat bandpass centered at 100.0 GHz with bandwidth 0.23 GHz. Figure~\ref{fig:res_NOinp_inp_d0s0} shows the residual maps obtained by preprocessing with STITCH (assuming noise parameters $f_\text{knee}~100\,\mathrm{mHz}$ and $\slopef=2$). The corresponding $BB$ residual spectra, together with those obtained for $\slopef=1$, are shown in figure~\ref{fig:res_spectra_d0}. Consistently with what we did in section~\ref{subsec:CMB}, we fit it for $r$ and $A_\text{lens}$ over the multipole ranges $2\leq\ell\leq29$ and $30\leq\ell\leq200$. The resulting best-fit values quantify the residual in terms of an effective $r$ and $A_\text{lens}$. We obtain $\{r=2.9\times10^{-5},\,A_\text{lens}=0.0\}$, over the low-multipole range, and $\{r=4.6\times10^{-7},\,A_\text{lens}=2.7\times10^{-6}\}$, over the high-multipole range. 

Although these residuals remain small compared to the input signal (roughly, four orders of magnitude), they require careful treatment to avoid biasing the recovered $B$ modes. As residuals might be projected through component separation, more detailed assessment of their impact, after component-separation, is left for future work.


\subsection{Interplay with sub-pixel effects}\label{subsec:subpix}
So far we have assumed that all TOD samples fall at the centers of their corresponding pixels. This isolates the effect of STITCH itself, but neglects sub-pixel effects arising from the finite pixelization of the sky. In realistic observations, detector pointings are distributed across each pixel, and replacing the continuous sky by its pixelized representation can introduce additional residuals, particularly in the presence of strong sky gradients.

We therefore compare the residuals introduced by STITCH with those arising from sub-pixel effects. In particular, we quantify the associated $E\to B$ leakage. We assume a sky containing CMB and Galactic foreground emission (PySM \texttt{d0s0} model), with the input $B$ modes removed. We use the same instrumental configuration introduced in section~\ref{subsec:noise}, with two pairs of orthogonal detectors\minor{, with a separation of roughly 50 arcmin along the scanning direction}. This configuration ensures that the reconstructed maps are not affected by $I\to P$ leakage, in absence of other systematic effects as for this work.

We simulate two sets of TOD, $\mathbf{d}$ and $\mathbf{d}^{\rm real}$, using pixel-centered and realistic pointings, respectively. Each TOD is then processed in chunks with circulant GLS, with and without STITCH preprocessing, resulting in four sets of maps: $\widehat{\mathbf{m}}_\text{ref}$, $\widehat{\mathbf{m}}_\text{STITCH}$, $\widehat{\mathbf{m}}^\text{real}_\text{ref}$, and $\widehat{\mathbf{m}}^\text{real}_\text{STITCH}$. We define the difference map
\begin{equation}
    \Delta\widehat{\mathbf{m}}_{\rm subpix} = \widehat{\mathbf{m}}^\text{real}_\text{ref} - \widehat{\mathbf{m}}_\text{ref}\,,
\end{equation}
which isolates the map-level difference associated with sub-pixel effects. We then use the $BB$ power spectrum of this difference map to quantify the corresponding $E\to B$ leakage.

\begin{figure}[!t]
    \centering
    \includegraphics[trim={0 0 0 6mm},clip]{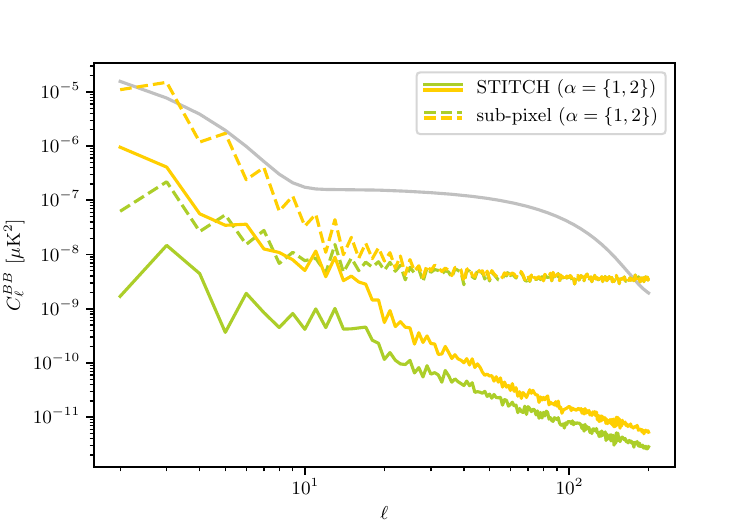}
    \caption{Angular $BB$ power spectra quantifying the $E\to B$ leakage introduced by STITCH (solid lines) and sub-pixel effects (dashed lines). Green lines are obtained with $\slopef=1$, while yellow lines with $\slopef=2$. In all cases, we assume CMB, dust and synchrotron in input. For reference, we also plot primordial $B$ modes with $r=10^{-3}$ (in gray).
    }
    \label{fig:inp_vs_subpix}
\end{figure}

\minor{Figure~\ref{fig:inp_vs_subpix} compares the $BB$ power spectra of $\Delta\widehat{\mathbf{m}}_{\rm subpix}$ and \minor{$\Delta\widehat{\mathbf{m}}_\text{STITCH}$, defined as in eq.~\eqref{eqn:deltastitch}}. For reference, we also show the primordial $B$-mode spectrum for $r=10^{-3}$. The STITCH residuals (solid lines) are consistent with those obtained in the foreground-only simulations shown in figure~\ref{fig:res_spectra_d0}. For both $\slopef=1$ and $\slopef=2$, they remain subdominant to the leakage induced by sub-pixel effects (dashed lines). These results suggest that, for the configuration considered, STITCH residuals are subdominant to those arising from sub-pixel effects, which may therefore represent a more significant source of bias. To further investigate these effects, we examine how they vary with the number of detectors.}

\begin{figure}[!t]
    \centering
    \includegraphics[trim={0 0 0 6mm},clip]{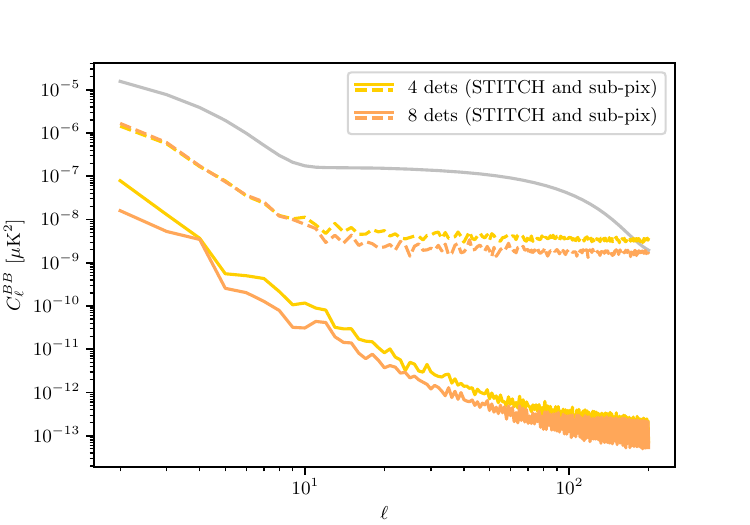}
    \caption{Dependence on the number of detectors of the residual angular $BB$ power spectra due to $E\to B$ leakage introduced by STITCH (solid lines) and sub-pixel effects (dashed lines). Yellow lines were obtained with two detector pairs, while orange lines with four. In all cases, we assume only CMB in input and $\slopef=2$. For reference, we also plot primordial $B$ modes with $r=10^{-3}$ (in gray).
}
    \label{fig:subpix_8}
\end{figure}

\paragraph{Dependence on the number of detectors}
We repeat the analysis with two additional detector pairs, offset by $\doubleoffsetarcmin$ arcmin along the $y$ axis of the focal plane, approximately orthogonal to the scanning direction. In this case, we use a CMB-only sky with no input $B$ modes and assume $\slopef=2$ for the preprocessing. We compute the $BB$ power spectra of $\Delta\widehat{\mathbf{m}}_{\rm subpix}$ and $\widehat{\mathbf{m}}_\text{STITCH}$ and compare them with the corresponding spectra obtained with two detector pairs, as shown in figure~\ref{fig:subpix_8}. 

Increasing the number of detectors suppresses the STITCH residual, while the effect of sub-pixel sampling depends on angular scale. At high multipoles, the sub-pixel residual is reduced as the number of detectors increases, whereas its large-scale component remains comparatively unchanged, {at around $10\%$ of the primordial $B$ mode power for $r=10^{-3}$}. This behavior can be understood as the combination of two effects.
\begin{itemize}[topsep=2mm,parsep=1mm,itemsep=1mm,leftmargin=6mm]
    \item[$\triangleright$] The GLS map-maker assumes that the signal is constant within each pixel. Signal variations on sub-pixel scales are therefore attributed to low frequency noise by the map-maker and weighted less in the solution, producing additional large-scale structure in the form of striping. This contribution is intrinsic to the pixelization of the map and is not eliminated by increasing the number of samples within each pixel.
    \item[$\triangleright$] Second, the signal measured by the TOD corresponds, on average, to the barycenter of the detector hits within each pixel rather than to the pixel center, introducing an effective pointing mismatch. Because the hit distribution varies from pixel to pixel, this contribution produces small-scale, approximately white residuals. Increasing the number of detectors increases the number of hits per pixel and brings the hit barycenter closer to the pixel center, thereby reducing this contribution. This contribution is white because nearly independent from pixel to pixel (GP).
\end{itemize}
These results indicate that sub-pixel effects can become an important source of bias for experiments targeting extremely faint primordial $B$ modes. Beam asymmetries, which are neglected in this work, can introduce additional residuals because a detector samples different signals when observing the same sky location with different orientations. A detailed treatment of these effects, including their interplay with map-making and scanning strategy, is left to future work.

\section{Discussion and conclusions}\label{sec:conclusions}
Time-correlated $1/f$ noise is a major challenge for large-scale CMB polarization measurements. Generalized least-squares (GLS) map-making provides an optimal solution for correlated Gaussian noise, but practical implementations rely on approximations to the time-domain noise covariance. A particularly convenient one is the circulant approximation, which allows for efficient FFT based solutions. When the TOD is divided into finite chunks, however, this approximation introduces artificial discontinuities at the chunk boundaries, producing large-scale residuals in the reconstructed maps.

Using simulations, we showed that STITCH substantially suppresses the large-scale residuals introduced by processing timelines in chunks. The recovered noise power spectra closely match those obtained when the circulant approximation is valid from the outset, showing that STITCH can recover the reference solution to good accuracy in the configurations considered. We further showed that the method introduces negligible bias into the reconstructed CMB signal and has only a small residual impact on the large-scale $BB$ power spectrum. These results indicate that STITCH can substantially reduce edge effects while keeping the induced map residuals small for the configurations tested.

We then used STITCH to investigate how temporal $1/f$ noise projects onto angular scales for three representative satellite scanning strategies. We characterized the resulting $BB$ noise spectra using an effective angular noise model and found that the angular knee scale, $\ell_\mathrm{knee}$, increases approximately linearly with the temporal knee frequency, $f_\mathrm{knee}$, for a given scanning strategy. This motivates a simple rescaling that captures the dependence on $f_\mathrm{knee}$ and allows spectra with different temporal knee frequencies to be described by a common effective model. {On the other hand, rescaling by the scan speed alone does not fully align the results across scanning strategies, indicating that the temporal-to-angular mapping also depends on scan geometry. The empirical relations derived here therefore characterize the specific scanning strategies studied rather than defining a universal conversion between temporal and angular scales. STITCH makes such investigations computationally more accessible by enabling chunked circulant GLS map-making at lower computational cost than approaches that explicitly handle the full time-domain covariance, such as full-Toeplitz GLS.}

{We also compared STITCH residuals with those arising from sub-pixel effects when realistic pointings are used instead of assuming observations fall at pixel centers, finding the former to be subdominant. Our results suggest that sub-pixel effects can be an important source of bias for experiments targeting extremely faint primordial $B$ modes: at large scales, they reach around $10\%$ of the primordial $B$-mode power in the CMB-only case, increasing by an order of magnitude when Galactic foregrounds are included. These effects could be mitigated using pixel-less map-making, \minor{generalizing \cite{artdeco}}, which we leave to future work. Such methods could also be extended to account for beam asymmetries, which are neglected here and can introduce additional residuals because detectors observe the same sky location at different orientations.}

Several simplifying assumptions were adopted throughout this work. We assumed piecewise stationary Gaussian noise and focused exclusively on the edge effects associated with the circulant approximation, without considering gaps in the time-ordered data. The simulations were also restricted to idealized detector configurations and symmetric beams. The statistical formulation of STITCH suggests possible extensions beyond the configurations studied here. In particular, flagged samples or data gaps can be treated as missing regions to be reconstructed from the surrounding observations. Assessing the performance of these extensions remains future work. We also assumed that the temporal noise parameters are known exactly. In realistic analyses, these quantities are themselves estimated from the data, and propagating the resulting uncertainties through the STITCH preprocessing will be an important step toward applying the method to observational data. We leave this investigation to future work.

The STITCH implementation developed in this work has been incorporated into the \texttt{BrahMap} framework, providing a practical route to applying circulant GLS map-making to chunked TOD. Overall, our results show that STITCH can mitigate the edge effects associated with the circulant approximation in the configurations tested. The empirical temporal-to-angular noise relations further illustrate how the impact of \(1/f\) noise depends on the scanning strategy, {while our comparison with sub-pixel effects highlights the importance of controlling other sources of residuals in experiments targeting primordial $B$ modes.} Together, these results provide a starting point for more comprehensive studies of correlated noise and map-making in future CMB polarization missions.

\acknowledgments
We thank the MaPLEs team of the LiteBIRD Collaboration, as well as Radek Stompor and Paolo Natoli, for useful discussions during the preparation of this manuscript. Kavli IPMU is supported by World Premier International Research Center Initiative (WPI), MEXT, Japan. This work has also received funding by the European Union’s Horizon
2020 research and innovation program under grant agreement No 101007633 CMB-Inflate. AA is supported by the InDark and LiteBIRD INFN projects. The authors gratefully acknowledge the computational and data resources provided by the Leibniz Supercomputing Centre (\url{www.lrz.de}).
During the preparation of this manuscript, the authors used OpenAI's ChatGPT to assist with improving the clarity, organization, and language of portions of the text. The AI tool was not used to generate scientific ideas, perform analyses, or interpret results.

\appendix

\section{Choice of method parameters}\label{subsec:params}
Throughout this work, we have adopted $n_\mathrm{ext}=2^{14}$, corresponding to $T_{\rm ext} \simeq 16\,$min, and $n_\text{bin}=8$ as the reference values for the STITCH parameters. In this section, we justify this choice by quantifying the impact of both parameters on the reconstructed maps and the overall computational cost.

To compare different choices of parameter methods, we simulate three sets of timelines: circulant noise-only ($f_{\rm knee}=100\,{\rm mHz}$ and $\slopef = 2$), CMB-only and \texttt{d0s0}-only. We produce the corresponding reference maps, $\widehat{\mathbf{m}}_\text{ref}$, by applying the circulant GLS over $\sim23$ day long segments. Then, we split the timelines into $\sim 1.2$ day long chunks and process them with STITCH for different parameter choices.

We define the relative polarization root mean square (RMS) of the map difference
\begin{equation}\label{eqn:RMS}
    \frac{\mathrm{RMS}(\Delta \widehat{\mathbf{m}})}{\mathrm{RMS}(\widehat{\mathbf{m}}_\text{ref})} = \sqrt{\frac{\sum (\Delta Q^2_p + \Delta U^2_p)}{\sum (Q^2_p + U^2_p)}}\,,
\end{equation}
where $\Delta\widehat{\mathbf{m}} \equiv \widehat{\mathbf{m}}_\text{STITCH} - \widehat{\mathbf{m}}_\text{ref}$ and the sums span the pixel indices, $p$. 
This metric quantifies the relative perturbation introduced by the inpainting procedure and is used throughout this section to compare different choices of $n_\mathrm{ext}$ and $n_\text{bin}$.

Table~\ref{tab:method_params} summarizes the results. Increasing the length of the synthetic extension, $T_{\rm ext}$, from 8 to 32 minutes reduces the residual noise, though all values obtained are negligible, {while having a negligible impact on the relative RMS for CMB and dust}. Conversely, varying the binning factor over the range considered has little effect on the reconstruction quality.

Although a longer synthetic extension increases the size of the Fourier transforms performed at each conjugate-gradient iteration, it may also improve the conditioning of the linear system and thereby reduce the number of iterations required for convergence. Our timing results suggest that these two effects compete for the CMB and \texttt{d0s0} cases, leading to a non-monotonic dependence of the runtime on $n_\mathrm{ext}$. Conversely, increasing $n_{\rm bin}$ reduces the computational time for both noise- and sky signal-only simulations. A detailed profiling of the solver is left for future work.

\begingroup
\renewcommand{\arraystretch}{1.35}
\begin{table}[!t]
 \centering
 \begin{tabular}{ccc||c|c|c||c|c|c}
 & & & \multicolumn{3}{c||}{relative RMS [$\times 10^{-3}$]} & \multicolumn{3}{c}{comp.\ time [min]} \\
 & & & ~~noise~~ & ~~CMB~~ & ~~\texttt{d0s0}~~ & ~noise~ & ~CMB~ & ~\texttt{d0s0}~ \\
 \hline
 \parbox[t]{0mm}{\multirow{3}{*}{\rotatebox[origin=c]{90}{$T_\mathrm{ext}$ [min]}}} & & 8.03 & 1.938 & 0.234 & 1.589 & 62 & 61 & 111\\
 & & \cellcolor{gray!15}16.1 & \cellcolor{gray!15}1.153 & \cellcolor{gray!15}0.234 & \cellcolor{gray!15}1.594 & \cellcolor{gray!15}70 & \cellcolor{gray!15}42 & \cellcolor{gray!15}74\\
 & & 32.1 & 0.853 & 0.235 & 1.596 & 82 & 48 & 78 \\
 \hline
 \parbox[t]{0mm}{\multirow{3}{*}{\rotatebox[origin=c]{90}{\phantom{[}$n_\text{bin}$\phantom{]}}}} & & 4 & 1.162 & 0.236 & 1.594 & 74 & 46 & 78 \\
 & & \cellcolor{gray!15}8 & \cellcolor{gray!15}1.153 & \cellcolor{gray!15}0.234 & \cellcolor{gray!15}1.594 & \cellcolor{gray!15}70 & \cellcolor{gray!15}42 & \cellcolor{gray!15}74 \\
 & & 16 & 1.157 & 0.234 & 1.593 & 42 & 40 & 72 
 \end{tabular}
 \caption{Relative map-difference RMS defined as in eq.~\eqref{eqn:RMS} for different choices of the method parameters, $T_\mathrm{ext}$ and $n_\text{bin}$, per component. The reference values adopted in this work are highlighted in light gray. The computational time includes both the STITCH preprocessing and GLS map-making steps. Note that different noise realizations can take more or less time to converge.}
 \label{tab:method_params}
\end{table}
\endgroup

\bibliographystyle{JHEP}
\bibliography{bibliography}

\end{document}